\documentclass[preprint,12pt,3p]{elsarticle}

\usepackage{amssymb}
\usepackage{amsmath}

\iftrue 
\newcommand{\todo}[1]{{\textcolor{red}{[[TODO: {#1}]]}}}
\newcommand{\commenttext}[1]{{\textcolor{red}{[[{#1}]]}}}
\newcommand{\commentfoot}[1]{\footnote{\textcolor{red}{\emph{Comment: #1}}}}
\newcommand{\topic}[1]{}
\else
\newcommand{\todo}[1]{}
\newcommand{\commenttext}[1]{}
\newcommand{\commentfoot}[1]{}
\newcommand{\topic}[1]{}
\fi

\iffalse

        \newcommand{\cutsubsectionup}{\vspace*{-0.09in}}

        \newcommand{\cutparagraphup}{\vspace*{-0.17in}}

\else 

        \newcommand{\cutsubsectionup}{}

        \newcommand{\cutparagraphup}{}

\fi

\journal{Computational Material Science}

\begin{document}

\begin{frontmatter}

\title{Improving Direct Physical Properties Prediction of Heterogeneous Materials from Imaging Data via Convolutional Neural Network and a Morphology-Aware Generative Model}

\author[label1]{Ruijin Cang\corref{cor2}}
\ead{cruijin@asu.edu}
\address[label1]{Mechanical Engineering, Arizona State University, Tempe}
\cortext[cor2]{Equal contribution}

\author[label2]{Hechao Li\corref{cor2}}
\ead{hechaoli@asu.edu}
\address[label2]{Material Science and Engineering, Arizona State University, Tempe}

\author[label1]{Hope Yao}
\ead{houpu.Yao@asu.edu}

\author[label2]{Yang Jiao}
\ead{yjiao13@asu.edu}

\author[label1]{Yi Ren\corref{cor1}}
\ead{yiren@asu.edu}
\ead[url]{designinformaticslab.github.io}
\cortext[cor1]{Corresponding author}

\begin{abstract}
Direct prediction of material properties from microstructures through statistical models has shown to be a potential approach to accelerating computational material design with large design spaces. However, statistical modeling of highly nonlinear mappings defined on high-dimensional microstructure spaces is known to be data-demanding. Thus, the added value of such predictive models diminishes in common cases where material samples (in forms of 2D or 3D microstructures) become costly to acquire either experimentally or computationally. To this end, we propose a generative machine learning model that creates an arbitrary amount of artificial material samples with negligible computation cost, when trained on only a limited amount of authentic samples. The key contribution of this work is the introduction of a morphology constraint to the training of the generative model, that enforces the resultant artificial material samples to have the same morphology distribution as the authentic ones. We show empirically that the proposed model creates artificial samples that better match with the authentic ones in material property distributions than those generated from a state-of-the-art Markov Random Field model, and thus is more effective at improving the prediction performance of a predictive structure-property model. 
\end{abstract}


\end{frontmatter}



\section{Introduction}
\label{sec:intro}
Direct prediction of material properties through predictive models has attracted interests from both material and data science communities. Predictive models have the potential to mimic highly nonlinear physics-based mappings, thus reducing dependencies on numerical simulations or experiments during material design, and enabling tractable discovery of novel yet complex material systems~\cite{jain2013commentary,hautier2010data,kondo2017microstructure}. Nonetheless, the construction of predictive models for nonlinear functions, such as material structure-property mappings, is known to be data-demanding, especially when the inputs, e.g., material microstructures represented as 2D or 3D images, are high-dimensional~\cite{huan2016polymer}. Thus, the added value of predictive models quickly diminishes as the acquisition cost increases for material samples. We investigate in this paper a computational approach to generate artificial material samples with negligible cost, by exploiting the fact that all samples within one material system share similar morphology. More concretely, we define morphology as a style vector quantified from a microstructure sample, and propose a generative model that learns from a small set of authentic samples, and creates an arbitrary amount of artificial samples that share the same distribution of morphologies as the authentic ones. 

The key contribution of the paper is the introduction of a morphology constraint on the generative model that significantly improves the morphological consistency between the artificial and authentic samples from benchmark generative models. 
To demonstrate the utility of the proposed model, we run a case study on the prediction of the Young's modulus, the diffusion coefficient and the permeability coefficient of sandstone microstructures. We show that the generated artificial samples from the proposed model can improve the prediction performance more effectively than those from a state-of-the-art Markov Random Field (MRF) model. 


As an overview, the proposed model follows the architecture of a variational autoencoder~\cite{kingma2013auto} that learns to encode material microstructures into a lower-dimensional latent space and to decode samples from the latent space back into microstructures. 
Both the encoder and the decoder are composed of feed-forward convolutional neural networks for extracting and generating local morphological patterns, and are jointly trained to minimize the discrepancy between the artificial and authentic samples.
The target morphology is quantified from the authentic samples by an auxiliary network. 
The idea of quantifying material morphology through a deep network is inspired by the style transfer technique originally developed for image synthesis~\cite{gatys2016image}. 


The rest of the paper is structured as follows: In Sec.~\ref{sec:background} we review related work on material representations and reconstruction, based on which we delineate the novelty of this paper. We then introduce background knowledge on variational autoencoder and style transfer. Sec.~\ref{sec:model} elaborates on the details of the proposed model. Sec.~\ref{sec:results} presents a case study on the prediction of sandstone properties, where we demonstrate the superior performance of the proposed model against the benchmarks, in both microstructure generation, and the resultant property prediction accuracy. In Sec.~\ref{sec:discussion} we summarize findings from the case study and propose potential future directions. Sec.~\ref{sec:conclusions} concludes the paper.     

\section{Background}
\label{sec:background}

\subsection{Data science challenges in computational material science}
Incorporating data science into material discovery~\cite{jain2013commentary} and design~\cite{curtarolo2013high} faces unique challenges with high dimensionality of material representations and the lack of material data due to high acquisition costs.
We review existing work that address these challenges to some extent. 

\cutparagraphup
\paragraph{Challenge 1: Mechanisms for understanding material representations} A common approach to addressing the issue of high dimensionality is to seek for a representation, i.e., an encoder-decoder pair, for a material system: The encoder transforms microstructures to their reduced representations, and the decoder generates (i.e., reconstructs) them back from their representations. A good encoder-decoder pair should both achieve significant dimension reduction, and good matching between the data distribution (i.e., the distribution of authentic samples) and the model distribution (i.e., the distribution defined by the decoder). This is often feasible for material systems with consistent and quantifiable morphologies among their samples, as reviewed below. 

Existing encoders for material representations can be categorized as physical and statistical, some of which have led to accelerated design of various material systems~\cite{torquato2013random, jiang2013efficient, grigoriu2003random, xu2014descriptor, yu2017characterization}. Among all, {\bf physical encoders} characterize microstructures using composition (e.g., the percentage of each material constituent)~\cite{broderick2008informatics,ashby2005materials}, dispersion (e.g., inclusions' spatial relation, pair correlation,the ranked neighbor distance ~\cite{steinzig1999probability, tewari2004nearest, rollett2007three,borbely2004three, pytz2004microstructure,scalon2003spatial}), and geometry features (e.g., the radius/size distribution, roundness, eccentricity, and aspect ratio of elements of the microstructure~\cite{rollett2007three, steinzig1999probability, torquato2013random, sundararaghavan2005classification, basanta2005using, holotescu2011prediction, klaysom2011effects, gruber2010misorientation}). 
Among {\bf statistical encoders} are the N-point correlation functions~\cite{liu2013computational,borbely2004three, torquato2013random, sundararaghavan2005classification, basanta2005using}. Torquato et al.~\cite{rintoul1997reconstruction,yeong1998reconstructing,torquato2013random} show that the microstructure of heterogeneous materials can be characterized statistically via various types of N-point correlation functions~\cite{okabe2005pore, hajizadeh2011multiple}. Similar descriptors include lineal path function~\cite{oren2003reconstruction} and statistics calculated based on the frequency domain using fast Fourier transformation~\cite{fullwood2008gradient,fullwood2008microstructure}.
Another type of statistical encoders are {\bf random fields}~\cite{roberts1997statistical,bostanabad2016stochastic,liu2015random}, which define joint probability functions on the space of microstructures. Typical probability models include Gaussian random fields~\cite{quiblier1984new, ostoja1998random, torquato2013random} which treats binary microstructure images as level sets, and Markov random fields, where each pixel of the microstructure is assumed to be drawn from a probability function conditioned on its neighbouring pixels~\cite{bostanabad2016stochastic}.

{\bf Decoding of representations}, i.e., generation of microstructure through existing physical and statistical representations, involves optimization in the microstructure space: For physical representations and N-point correlation functions, a microstructure is searched to minimize its difference from the target descriptors. For random fields, the generation can be done by maximizing the joint probability through Markov chain Monte Carlo simulations~\cite{bostanabad2016stochastic,liu2015random}. While it is shown that material generation through these representations is feasible~\cite{borbely2004three, basanta2005using,torquato2013random,xu2014descriptor}, the computational costs for the optimization through gradient~\cite{fullwood2008gradient,xu2014descriptor} and non-gradient~\cite{yeong1998reconstructing,jiao2008modeling,jiao2009superior,karsanina2015universal} methods are often high.

\begin{figure}[htp]
\centering
\includegraphics[width=0.8\textwidth]{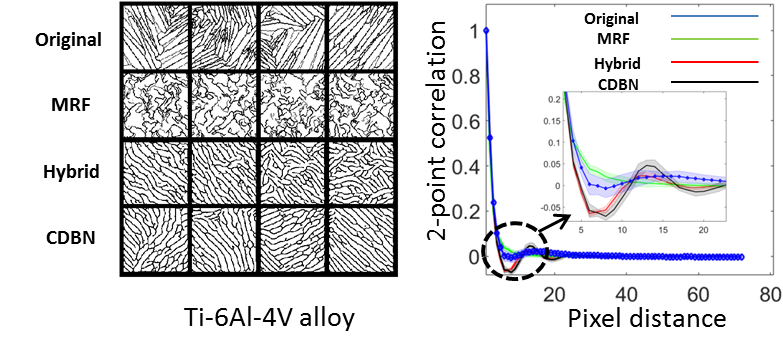}
\caption{Comparison of two-point correlation functions among four sets of images. From left to right: Authentic microstructure samples, samples generated by a Markov random field model~\cite{bostanabad2016stochastic}, samples generated by a hybrid model with deep belief network and Markov random field~\cite{cang2017scalable}, and samples generated by a deep belief network~\cite{cang2017microstructure}. Better matching in the discretized 2-point correlation space does not indicate better microstructure generations.}
\label{fig:point_corr}
\end{figure}

In addition to the difficulties in decoding, the existing encoders are not universally applicable, especially to material systems with complex morphology. More specifically, matching in the representation space does not guarantee the match in the microstructure space. An example can be found in Fig.~\ref{fig:point_corr}, where we compare two-point correlation functions of Ti64 alloy samples and three sets of artificial images (see details from \cite{bostanabad2016stochastic,cang2017microstructure,cang2017scalable}). The visually more plausible set has worse match to the target with respect to the Euclidean distance in the discretized 2-point correlation space. 

These existing difficulties leads to the need for new mechanisms to define material representations. We propose four metrics for evaluating the utility of a material representation: interpretability, dimensionality, expressiveness, and generation cost: 
Physical descriptors and correlation functions are designed to be interpretable and relatively low dimensional, yet may not be expressive enough to capture complex morphologies and requires optimization during generation; random fields are relatively expressive (and some permit fast generation~\cite{bostanabad2016stochastic}), but are often high-dimensional and less interpretable. Both categories of representations are material specific, i.e., new representations need to be manually identified for new material systems. Cang et al.~\cite{cang2017microstructure} proposed to learn statistical generative models from microstructures to automatically derive expressive, low dimensional representations that enables fast microstructure generation. They showed that a particular type of generative model, called Convolutional Deep Belief Network (CDBN)~\cite{lee2009convolutional}, can produce reasonable microstructures for material systems with complex morphologies, by extracting morphology patterns at different length scales from samples, and decode an arrangement of these patterns (the hidden activation of the network) back into a microstructure. 
Nonetheless, CDBNs are trained layer-by-layer, and thus require additional material-dependent parameter tuning to achieve plausible generations.

Different from \cite{cang2017microstructure}, this paper proposes a model that directly enforces the matching of the artificial microstructures to the authentic ones, thus avoiding additional parameter tuning. Our model is also fully differentiable, making it much easier to train than CDBNs and scalable to deep architectures. 
Lastly, the introduction of the morphology constraint further refines the artificial microstructures under a small amount of training data. With all these improvements, the proposed model is generally applicable to the extraction of low dimensional material representations, and is expressive for generating microstructures with morphologies of decent complexity. 

\cutparagraphup
\paragraph{Challenge 2: Effective material data acquisition methods} Unlike scenarios in contemporary machine learning where a large quantity of data is available, in computational material design tasks we may only have a limited amount of microstructure and property samples, and the acquisition cost for additional samples is usually high. This challenge calls for the incorporation of active learning methods into the design. The key idea of active learning is to minimize the data acquisition cost for learning a model (e.g., generative models for capturing microstructure distributions, or discriminative models for predicting process-structure-property mappings), by optimally controlling the balance between data exploitation and exploration~\cite{tong2001support1,settles2010active}. In computational material science, such techniques have been adopted to accelerate the process of material discovery for desired properties~\cite{ling2017high,lookman2016perspective}. 
While this paper does not focus directly on active learning, we demonstrate that the quality of the microstructure data, along with its quantity, can significantly influence the prediction performance of a statistical structure-property mapping, and thus justify the value of the proposed generative model.


\subsection{Preliminaries on predictive and generative models}
This paper will involve three networks: The generative network for creating artificial material microstructures, the auxiliary convolutional network for quantifying material morphologies, and another convolutional network for structure-property prediction. We provide technical backgrounds for these models below.

\begin{figure}[htp]
\centering
\includegraphics[width=0.8\textwidth]{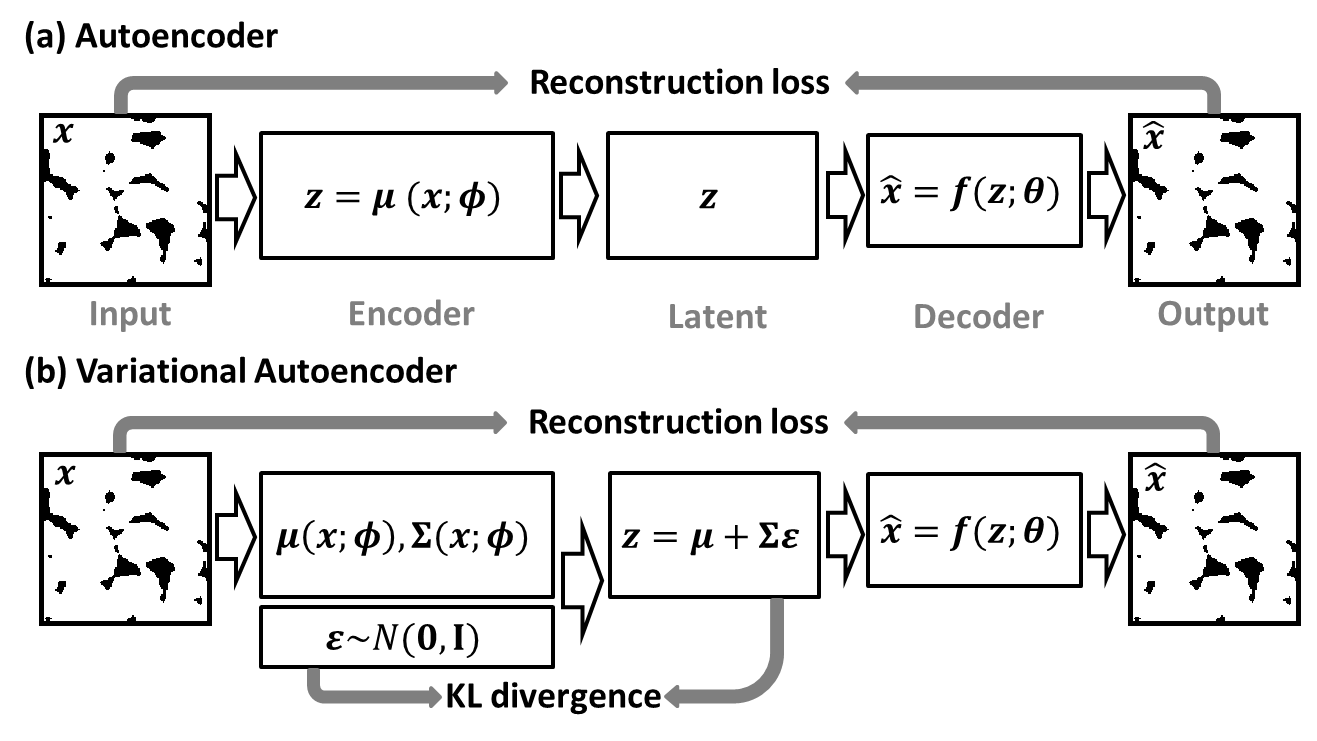}
\caption{(a) An autoencoder converts inputs to a latent space and reconstructs outputs from samples in the latent space. It is trained to minimize the reconstruction error. (b) A variational autoencoder converts inputs to a distribution in the latent space. It is trained to minimize both the reconstruction error and the Kullback-Leibler (KL) divergence between the latent distribution and a sample distribution (often standard normal).}
\label{fig:AE}
\end{figure}

\cutparagraphup
\paragraph{Generation through Variational Autoencoder}
\label{para:VAE}
Variational autoencoders (VAE)~\cite{kingma2013auto} are extensions of autoencoders~\cite{bengio2009learning}. An autoencoder (Fig.~\ref{fig:AE}a) is composed by two parts: an encoder ${\bf z}=f({\bf x})$ converts the input $\textbf{x}$ to a hidden vector $\textbf{z}$, and a decoder produces a reconstruction $\hat{\textbf{x}} = g(\textbf{z})$. An AE is trained to minimize the discrepancy between inputs $\textbf{x}$ and their corresponding reconstructions $\hat{\textbf{x}}$. 
Variants of autoencoders (e.g., Sparse~\cite{ng2011sparse}, denoising~\cite{vincent2008extracting}, and contractive~\cite{rifai2011contractive}) have been proposed to improve the learning of more concise representations from high-dimensional input data,
and are widely used for data compression~\cite{le2015tutorial}, network pre-training~\cite{bengio2007greedy}, and feature extraction~\cite{xing2015stacked}. However, conventional autoencoders are prone to generating implausible new outputs, see Fig.~\ref{fig:gen_compare}b for examples, as they do not attempt to match the model distribution (the distribution of images generated from the decoder) with the data distribution (the distribution of input images). VAE (Fig.~\ref{fig:AE}b) was introduced to address this issue~\cite{kingma2013auto}, by constraining the distribution of the latent variables $\textbf{z}$ encoded from the input data to that used for output generation, thus indirectly forcing the match between the distributions of the model outputs and the data.

\cutparagraphup
\paragraph{The VAE model} We start with the decoder, which defines the output distribution, $p_{\boldsymbol{\theta}} (\textbf{x}|\textbf{z})$, conditioned on the latent vector $\textbf{z}$, and is parameterized by $\boldsymbol{\theta}$.  Let the input data be $\mathcal{X}=\{\textbf{x}_i\}^N_{i=1}$, which defines a data distribution $p(\textbf{x})$, and let $p(\textbf{z})$ be the pre-specified sampling distribution in the latent space. The goal is to train a model that matches the marginal distribution $p_{\boldsymbol{\theta}}(\textbf{x})$ to the data distribution $p(\textbf{x})$. In the following, we show that this is indirectly achieved by matching the posterior $p_{\boldsymbol{\theta}}(\textbf{z}|\textbf{x})$ to $p(\textbf{z})$.

First, the marginal is defined as
\begin{equation}
    p_{\boldsymbol{\theta}}(\textbf{x}) = \int_{\textbf{z}}p_{\boldsymbol{\theta}} (\textbf{x}|\textbf{z})p(\textbf{z})d\textbf{z} = E_{\textbf{z}\sim p(\textbf{z})} p_{\boldsymbol{\theta}}(\textbf{x}|\textbf{z}).
    \label{eq:px}
\end{equation}
Matching the model and data distributions requires maximizing the likelihood $\Pi_{\textbf{x}\in \mathcal{X}}p_{\boldsymbol{\theta}}(\textbf{x})$ with respect to $\boldsymbol{\theta}$. This maximization problem, however, is often intractable due to the numerical integral and the complexity of the decoder network. Nonetheless, it is noted that for a given $\textbf{x}$, most samples in the latent space will have $p_{\boldsymbol{\theta}} (\textbf{x}|\textbf{z}) \approx 0$, as these latent samples do not generate outputs close to $\textbf{x}$. This leads to the idea of introducing another distribution $q_{\boldsymbol{\phi}}(\textbf{z}|\textbf{x})$ (the encoder) that takes an $\textbf{x}$ and outputs $\textbf{z}$ values that are likely to reproduce $\textbf{x}$. Ideally, the space of $\textbf{z}$ that are likely under $q$ will be much smaller than that under $p(\textbf{z})$,  making the computation of $E_{\textbf{z}\sim q} p_{\boldsymbol{\theta}} (\textbf{x}|\textbf{z})$ relatively cheap. However, $E_{\textbf{z}\sim q} p_{\boldsymbol{\theta}} (\textbf{x}|\textbf{z})$ is not the same as $p(\textbf{x})$, as $q_{\boldsymbol{\phi}}(\textbf{z}|\textbf{x})$ does not necessarily match to $p(\textbf{z})$. The relation between the two is derived below. 


We start by deriving the Kullback-Leibler divergence between the encoded latent distribution $q_{\boldsymbol{\phi}}(\textbf{z}|\textbf{x})$ and the posterior $p_{\boldsymbol{\theta}}(\textbf{z}|\textbf{x})$: 
\begin{equation}
\begin{aligned}
    D_{KL}\left(q_{\boldsymbol{\phi}}(\textbf{z}|\textbf{x})||p_{\boldsymbol{\theta}}(\textbf{z}|\textbf{x})\right) & = \int q_{\boldsymbol{\phi}}(\textbf{z}|\textbf{x})\log\frac{q_{\boldsymbol{\phi}}(\textbf{z}|\textbf{x})}{p_{\boldsymbol{\theta}}(\textbf{z}|\textbf{x})}d\textbf{z}\\
    & =-\mathbb{E}_{\textbf{z} \sim q}[\log p(\textbf{z},\textbf{x})]-\mathbb{E}_{\textbf{z} \sim q}[-\log q_{\boldsymbol{\phi}}(\textbf{z}|\textbf{x})] + \log p(\textbf{x}).
\end{aligned}
\label{eq:KL}
\end{equation}
By rearranging Eq.~\eqref{eq:KL}, we get
\begin{equation}
    \log p(\textbf{x}) - D_{\text{KL}}\left(q_{\boldsymbol{\phi}}(\textbf{z}|\textbf{x})||p_{\boldsymbol{\theta}}(\textbf{z}|\textbf{x})\right) = \mathbb{E}_{\textbf{z} \sim q}[\log p_{\boldsymbol{\theta}}(\textbf{x}|\textbf{z})]
    -D_{\text{KL}}\left(q_{\boldsymbol{\phi}}(\textbf{z}|\textbf{x})||p(\textbf{z})\right).
\label{eq:vae}
\end{equation}
Note that the right hand side of Eq.~\eqref{eq:vae} can be maximized through stochastic gradient descent: The first term $\mathbb{E}_{\textbf{z} \sim q}[\log p_{\boldsymbol{\theta}(\textbf{x}|\textbf{z})}]$ measures the reconstruction error, i.e., the expected difference between an input $\textbf{x}$ and its reconstructions drawn from the decoder conditioned on the latent variables, which are further drawn from the encoder conditioned on the input. The second term $D_{\text{KL}}(q_{\boldsymbol{\phi}}(\textbf{z}|\textbf{x})||p(\textbf{z}))$ measures the difference between the encoded distribution of $\textbf{z}$ and the modeled latent distribution $p(\textbf{z})$. Since the encoding and decoding processes involved in these terms are modeled as feedforward networks, the gradients are readily available for backpropagation. Specifically, we model $q_{\boldsymbol{\phi}}(\textbf{z}|\textbf{x}) = \mathcal{N}(\textbf{z}|\boldsymbol{\mu}(\textbf{x};\boldsymbol{\phi}),\boldsymbol{\Sigma}^2(\textbf{x};\boldsymbol{\phi}))$ as a normal distribution parameterized by the mean $\boldsymbol{\mu}(\textbf{x};\boldsymbol{\phi})$ and the diagonal variance matrix $\boldsymbol{\Sigma}^2(\textbf{x};\boldsymbol{\phi})$. $\boldsymbol{\mu}$ and $\boldsymbol{\Sigma}$ are the outputs of the encoder network. We similarly model $p_{\boldsymbol{\theta}(\textbf{x}|\textbf{z})} = \mathcal{N}(\textbf{x}|\textbf{f}(\textbf{z};\boldsymbol{\theta}),\sigma^2\textbf{I})$ with mean $\textbf{f}(\textbf{z};\boldsymbol{\theta})$ and variance $\sigma^2$. The function $\textbf{f}(\textbf{z};\boldsymbol{\theta})$ is the decoder network; $\sigma$ determines the importance of the reconstruction of $\textbf{x}$ during the training of a generative model, and is set to $1$ in the proposed model. 
The prior of the latent distribution, $p(\textbf{z})$, is often assumed to be standard normal. This is because the transformation from this simple distribution to the potentially highly nonlinear data distribution $p(\textbf{x})$ can often be achieved by a sufficiently deep decoder. 


The left hand side of Eq.~\eqref{eq:vae} contains the objective $\log p_{\boldsymbol{\theta}}(\textbf{x})$ that we want to maximize, and the KL-divergence $D_{KL}(q_{\boldsymbol{\phi}}(\textbf{z}|\textbf{x})||p_{\boldsymbol{\theta}}(\textbf{z}|\textbf{x}))>0$ that should ideally reach 0. Thus, minimizing 
\begin{equation}
    L(\theta,\phi,\textbf{x}) = -\mathbb{E}_{\textbf{z} \sim q}[\log p_{\boldsymbol{\theta}}(\textbf{x}|\textbf{z})]
    +D_{\text{KL}}(q_{\boldsymbol{\phi}}(\textbf{z}|\textbf{x})||p(\textbf{z}))
\end{equation}
will maximize a lower bound of $p(\textbf{x})$.

\begin{figure}[htp]
\centering
\includegraphics[width=0.8\textwidth]{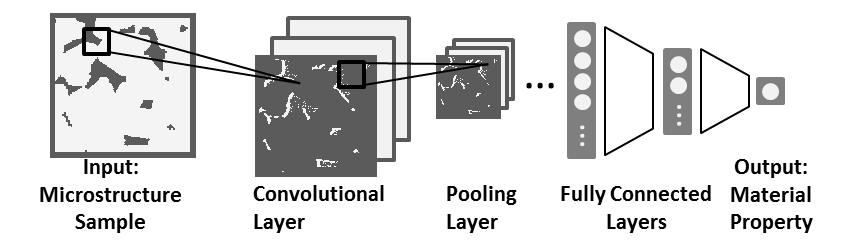}
\caption{Schematic plot for a CNN}
\label{fig:CNN}
\end{figure}

\cutparagraphup
\paragraph{Prediction through Convolutional Neural Network (CNN)}
We build a predictive structure-property model using CNN (Sec.~\ref{sec:model}). A CNN is often composed of one or more convolutional layers followed by fully connected layers. 
A convolutional layer consists of a set of image filters, each as a set of network edges connecting the input image to a hidden channel, see Fig.~\ref{fig:CNN}. Scanning the input image with a filter requires the network weights to be shared, thus reducing the total number of weights to be trained. A max-pooling layer further reduces the dimension of a hidden layer~\cite{szegedy2015going}.

\cutparagraphup
\paragraph{Style transfer (ST)}
\label{para:style_transfer}
ST was first introduced to integrate the content of one image with the style of another~\cite{gatys2016image}. To perform ST, a ``style vector" is first computed for an image, that consists of the variance-covariance matrices of the hidden states of a pre-trained CNN (we use a VGG Net~\cite{simonyan2014very} in this paper) activated by the input image. A new image can then be created by minimizing the difference between its style vector and the target one. One can also preserve the image content, i.e., the hidden states of the deepest layers, by adding an additional loss on the content difference.   



As will be discussed in Sec.~\ref{sec:model}, the proposed method extends ST to a generative setting, by directly adding a style (morphology) penalty to the training loss of a VAE. There are two major difference in comparison: First, the standard ST generates images by minimizing the style loss with respect to a high-dimensional image space, while our model is learned to directly generate style-consistent images, thus is significantly faster than standard ST; second, the standard ST uses a single style vector as a target, while our model can encode a distribution of styles in its latent space, and thus is more expressive as a generator. 



\section{Proposed Models}
\label{sec:model}

\subsection{Network specifications}

\begin{figure}[htp]
\centering
\includegraphics[width=0.8\textwidth]{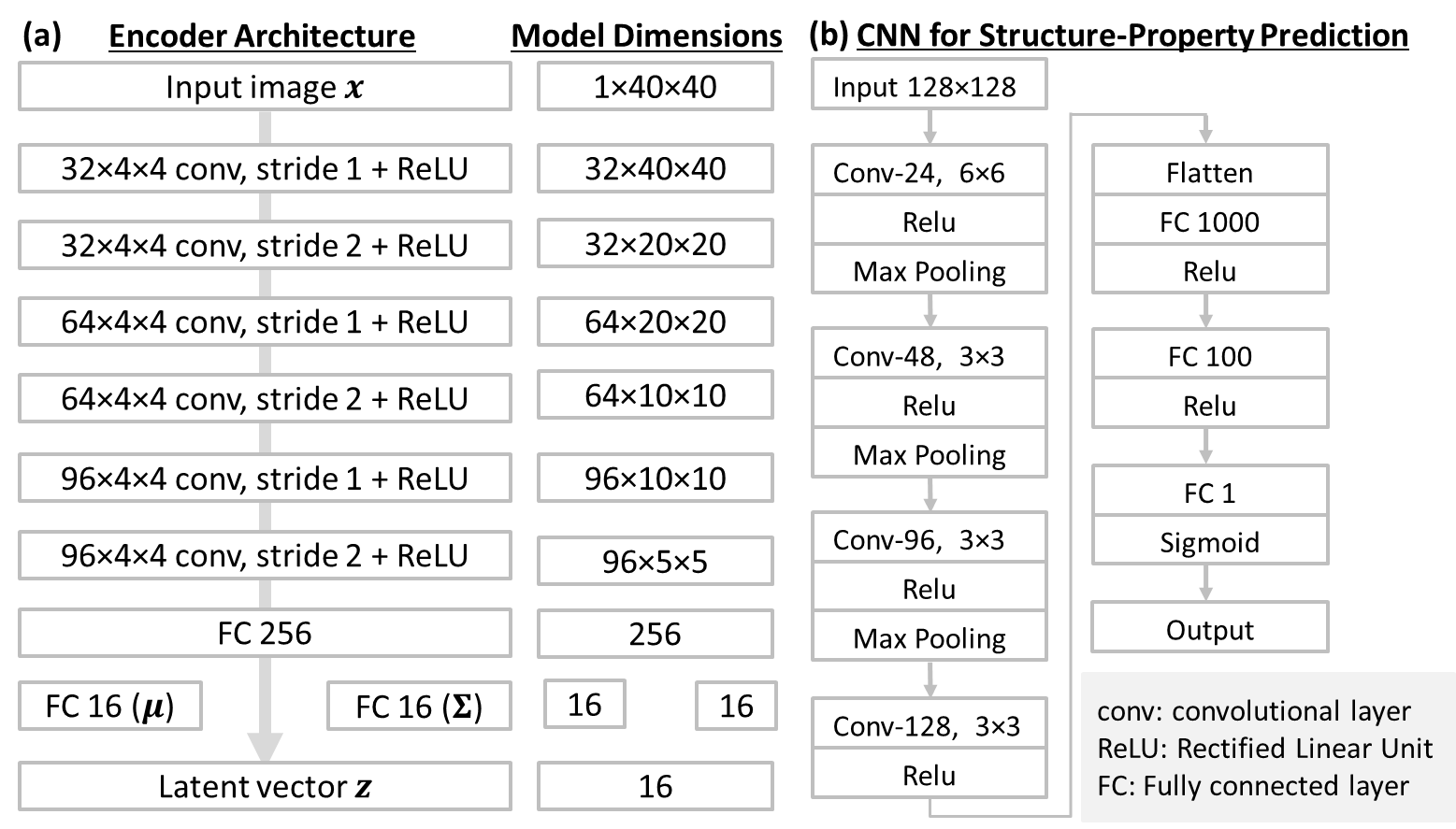}
\caption{(a) Encoder architecture in the VAE, decoder is mirrored from the encoder (b) CNN architecture}
\label{fig:VAE+CNN}
\end{figure}

\paragraph{The generative model}
The proposed generative model is composed of two networks, a VAE for image encoding and decoding, and an auxiliary network for computing the style vector ($\textbf{G}$). See model summary in Fig.~\ref{fig:VAE+CNN}a. For the VAE, its encoder has convolutional layers of a fixed filter size ($4\times4$), max-pooling layers after each other convolutional layers, and two fully-connected layers of sizes 256 and 16, respectively. 
Each pooling layer is of stride 2, i.e., it is composed with filters of size $2\times 2$ applied to the hidden layers, downsampling these layers by a factor of 2 along both width and height. Together, the encoder reduces the dimension of the input from $128\times128$ to $16\times1$. The architecture of the decoder is symmetric to that of the encoder. 

\cutparagraphup
\paragraph{The predictive model}
A CNN is built to predict material properties from microstructures. A schematic of the predictive model is in Fig.~\ref{fig:VAE+CNN}b. We use four convolutional layers with filter size $3\times 3$,
and max-pooling layers with stride 2. 

\begin{figure}[htp]
\centering
\includegraphics[width=0.8\textwidth]{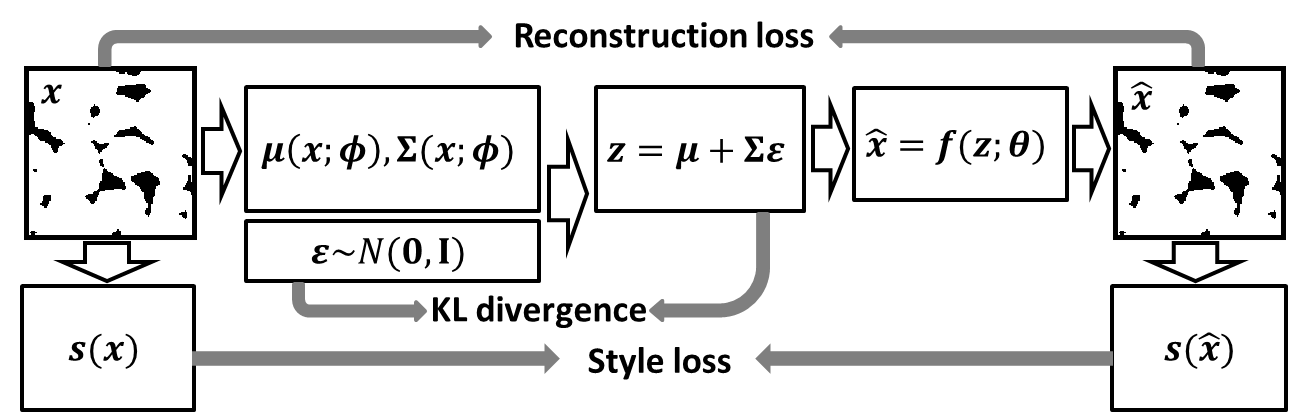}
\caption{Schematic of the proposed morphology-aware VAE}
\label{fig:VAE+morhoplogy}
\end{figure}

\cutparagraphup
\paragraph{The morphology style model}
To acquire a rich set of morphology patterns and make this acquisition process general, we propose to use a VGG network~\cite{Simonyan14c} as the morphology model. A VGG is trained on over a million images~\cite{imagenet_cvpr09} to learn a rich set of feature representations. The schematic for the proposed morphology-aware VAE is shown as Fig.~\ref{fig:VAE+morhoplogy}.

\subsection{Model training}

\paragraph{The loss function of the generative model}
Training the generative model involves minimizing a loss function with four components: \begin{equation}
    L({\boldsymbol{\theta}},{\boldsymbol{\phi}};\mathcal{X}) = L_{\text{RECON}} + L_{\text{KL}} + L_{\text{ST}} + L_{\text{MC}},
    \label{eq:obj}
\end{equation}
where $L_{\text{RECON}}+L_{\text{KL}}$ are the standard reconstruction and KL divergence losses for VAE (see Sec.~\ref{para:VAE}), $L_{\text{ST}}$ is the style loss, and $L_{\text{MC}}$ is an additional loss to prevent mode collapse (introduced below). 

The reconstruction loss 
\begin{equation}
\label{eq:recon}
L_{\text{RECON}}=-\sum_{\textbf{x}_i\in \mathcal{X}} \mathbb{E}_{q_{\boldsymbol{\phi}}(\textbf{z}|\textbf{x}_i)}[\log p_{\boldsymbol{\theta}}(\textbf{x}_i|\textbf{z})]
\end{equation}
measures the reconstruction error of all pairs of authentic microstructures ($\textbf{x}_i\in \mathcal{X}$) and their reconstructions. 
The KL divergence loss
\begin{equation}
\label{eq:kl}
L_{\text{KL}}=\sum_{\textbf{x}_i\in \mathcal{X}} D_{KL}(q_{\phi}(\textbf{z}|\textbf{x}_i)|| p(\textbf{z}))
\end{equation}
measures the difference between the encoder distribution $q_{\phi}(\textbf{z}|\textbf{x}_i)$ and the prior $p(\textbf{z})$. 
The style loss
\begin{equation}
L_{\text{ST}}=\sum_{s=1}^N\sum_{p=1}^N\sum_{l} ||\textbf{G}^{s,l}-\textbf{G}^{p,l}||^2_{F}
\end{equation}
measures the total style difference between two sets of $N$ images with indices $s$ and $p$, representing the generated and the authentic ones, respectively. Given a morphology style model, the Gram matrix $\textbf{G}^{s,l}$ of image $s$ at the $l$th hidden layer is defined through elements $G_{i,j}^{s,l}=\sum_k f_{i,k}^{s,l}f_{j,k}^{s,l}$, where feature element $f_{i,k}^{s,l}$ represents the hidden activation corresponding to the $i$th convolutional filter at spatial location $k$. $||\cdot||_F$ is the Frobenius normal. We note that the definition of feature maps will affect the outcome of style transfer. In general, deeper layers capture styles at larger length scales. We use the first four layers of a standard VGG net. This definition of features is empirical.

Lastly, the mode collapse loss is introduced to prevent the model from producing clustered samples~\cite{zhao2016energy}, by forcing the generated samples to be different in terms of their activations in the style network. The loss is defined as:
\begin{equation}
L_{\text{MC}}=\frac{2}{N(N-1)}\sum_{i=1}^N\sum_{j=i+1}^N(\frac{\textbf{s}_i^T \textbf{s}_j}{\|\textbf{s}_i\|\|\textbf{s}_j\|})^2,
\end{equation}
where the style vector $\textbf{s}$ contains the concatenated and vectorized Gram matrices as defined above. 

\begin{figure}[htp]
\centering
\includegraphics[width=16cm]{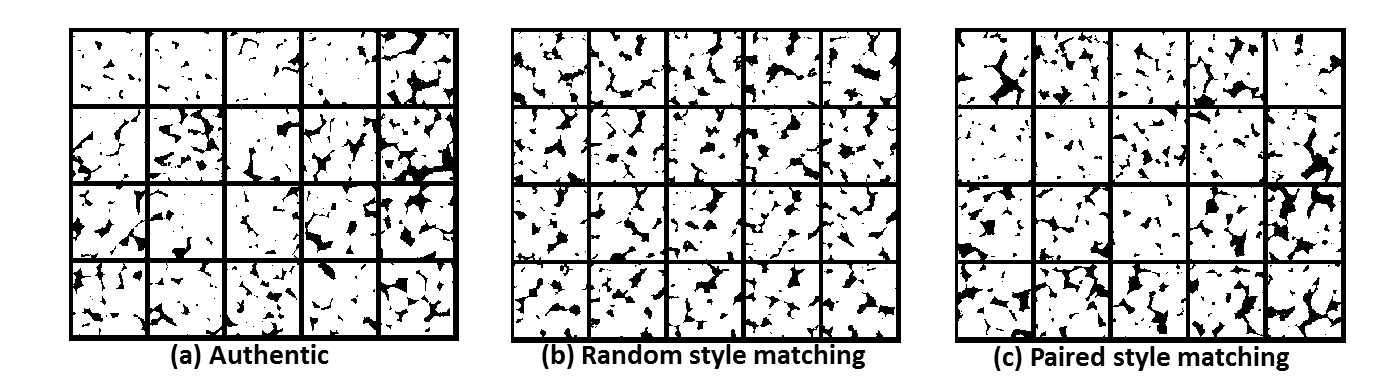}
\caption{(a) Samples of authentic microstructures (b) Artificial microstructures when random samples are drawn from a standard normal distribution during the training (c) Artificial microstructures when random samples are drawn to pair with the authentic ones}
\label{fig:proposed_model_comparison}
\end{figure}

\cutparagraphup
\paragraph{The training of the generative model}
The generative model is trained to minimize the loss defined in Eq.~\eqref{eq:obj} with respect to model parameters $\boldsymbol{\theta}$ and $\boldsymbol{\phi}$ through stochastic gradient descent. To approximate the gradient of the loss, $\nabla_{\boldsymbol{\theta},\boldsymbol{\phi}} (L_{\text{RECON}} + L_{\text{KL}} + L_{\text{ST}} + L_{\text{MC}})$, we randomly sample 20 authentic material microstructures from $\mathcal{X}$ and generate another 20 artificial microstructures in each training iteration. The artificial microstructures are needed to calculate $\nabla_{\boldsymbol{\theta},\boldsymbol{\phi}} (L_{\text{ST}} + L_{\text{MC}})$.
To prevent neighbouring latent samples to be matched to drastically different style (morphology) targets, we propose to create artificial microstructures by the following procedure: We take each authentic sample $\textbf{x}$ and pass it through the encoder to get $\boldsymbol{\mu}(\textbf{x};\boldsymbol{\phi})$ and $\boldsymbol{\Sigma}(\textbf{x};\boldsymbol{\phi})$. We then draw a random sample $\textbf{z}$ from the latent space following the normal distribution $\mathcal{N}(\textbf{z}| \boldsymbol{\mu}(\textbf{x};\boldsymbol{\phi}),\boldsymbol{\Sigma}(\textbf{x};\boldsymbol{\phi}))$, and derive its decoded microstructure $\textbf{x}' = f(\textbf{z};\boldsymbol{\theta})$ to pair with $\textbf{x}$. By matching the style of $\textbf{x}$ and $\textbf{x}'$, we ensure that samples close to each other in the latent space will have similar morphologies. We demonstrate the necessity of this treatment in Fig.~\ref{fig:proposed_model_comparison}, where we show the generation results from an alternative method where microstructures from a standard normal distribution are randomly paired with the authentic ones for style loss calculation, leading to averaged morphology across all random generations (Fig.~\ref{fig:proposed_model_comparison}b). 
We use Adam optimizer~\cite{kingma2014adam} for training, with 200K iterations and a learning rate of 0.001. 

\cutparagraphup
\paragraph{The training of the predictive model}
We divide the raw data into training, validation, and test sets. The loss function for training the predictive structure-property model is the mean square error of the predicted properties of the training data. Prediction performance of the model on the validation set is monitored to prevent the model from overfitting. The performance of the model is then measured by R-square value of the test data. We use Adam for training with a learning rate of 0.001.  



\section{Case Study and Results}
\label{sec:results}
In this section, we demonstrate key contributions of the proposed model through a case study: (1) By incorporating the style loss, the proposed generative model creates material microstructures with better visual and statistical similarity to the authentic ones than a state-of-the-art Markov random field (MRF) approach; and (2) we can improve the prediction of material properties more effectively by using the artificial data generated from the proposed model than by those generated from a MRF. The case study uses a set of sandstone microstructures and their properties, including Young's modulus, diffusivity, and fluid permeability. Details of this material system are introduced as follows.

\subsection{A case study on the sandstone system}
\label{sec:sandstone}
Sandstones are a class of important geological porous
materials in petroleum engineering. In the simplest form, a
mono-mineral sandstone is composed of a solid (rock) phase and a
pore (void) phase, both of which are typically percolating.
Extensive research work has been carried out to model the
microstructure and physical properties of such porous materials~\cite{sahimi2011flow, radlinski2004angstrom, milliken2000brittle,
blair1996using, coker1996morphology, antonellini1994petrophysical,
al2005extraction, appoloni2002characterization, li2017accurate,
hajizadeh2011multiple}. For example, a fast and independent
architecture of artificial neural network has been developed for
accurately predicting fluid permeability~\cite{tahmasebi2012fast}.

We will focus on three different material properties: Young's
modulus (stiffness) $E$, diffusivity $D$ and fluid permeability
$k$. These properties are respectively sensitive to distinct
microstructural features of the materials. In particular, $E$ is
mainly determined by the morphology and volume fraction of the
rock phase. The diffusivity $D$ is most sensitive to the local
pore size distribution. The fluid permeability $k$ depends on the
degree of connectivity of the pore phase as well as the pore/rock
interface morphology. Thus, successfully improving the prediction
accuracy for all three material properties poses a stringent test
for our morphology-aware generative model for effectively
populating the microstructure sample space.

\cutparagraphup
\paragraph{Physics-based material property calculations}
The material properties used to train the network model are first
computed using the effective medium theory~\cite{torquato2013random}. The Young's modulus can be written as
\begin{equation}
E = \frac{9 K_e G_e}{3K_e + G_e},
\end{equation}
where $K_e$ and $G_e$ are respectively the bulk and shear modulus
of the materials. We use analytical approximations of $K_e$ and
$G_e$ obtained by truncating the associated strong-contrast
expansions after the third order term~\cite{torquato1997exact},
which are respectively given by
\begin{equation}
\label{bulk} \phi_2 \frac{\kappa_{21}}{\kappa_{e1}} = 1 -
\frac{(d+2)(d-1)G_1\kappa_{21}\mu_{21}}{d(K_1 + 2G_1)} \phi_1 \chi
\end{equation}
and
\begin{equation}
\label{shear} \phi_2 \frac{\mu_{21}}{\mu_{e1}} = 1-
\frac{2G_1\kappa_{21}\mu_{21}}{d(K_1 + G_1)}\phi_1 \chi -
\frac{2G_1\kappa_{21}\mu_{21}}{d(K_1 + 2G_1)}\phi_1 \chi -
\frac{1}{2d}[\frac{dK_1 + (d-2)G_1}{K_1 + 2G_1}]^2\mu_{21}^2\phi_1
\eta_2,
\end{equation}
where $\phi_1$ and $\phi_2$ are respectively the volume fraction of
the rock phase and void phase (i.e., porosity); $d$ is the spatial
dimension of the material system; $K_p$ and $G_p$ are respectively
the bulk and shear modulus of phase $p$; and the scalar parameters
$\kappa_{pq}$ and $\mu_{pq}$ ($p, q = 1, 2, e$) are respectively
the bulk and shear modulus polarizability, i.e,
\begin{equation}
\kappa_{pq} = \frac{K_p - K_q}{K_q + \frac{2(d-1)}{d}G_q}
\end{equation}
and
\begin{equation}
\mu_{pq} = \frac{G_p - G_q}{G_q}\frac{1}{1+\frac{\frac{d}{2}K_q +
\frac{(d+1)(d-2)}{d}G_1}{K_q + 2G_q}}.
\end{equation}
The quantities $\chi$ and $\eta_2$ are the microstructural
parameters associated with the void (pore) phase involving the
three-point correlation functions $S_3$ and two-point correlation
functions $S_2$ of the pore phase, i.e.,
\begin{equation}
\label{chi} \chi=
\frac{9}{2\phi_1\phi_2}\int_0^{\infty}\frac{dr}{r}\int_0^{\infty}\frac{ds}{s}\int_{-1}^{1}d(\cos\theta)P_2(\cos\theta)[S_3(r,
s, t)-\frac{S_2(r)S_2(t)}{\phi_2}]
\end{equation}
and
\begin{equation}
\eta_2 = \frac{5}{21}\chi +
\frac{150}{7\phi_1\phi_2}\int_0^{\infty}\frac{dr}{r}\int_0^{\infty}\frac{ds}{s}\int_{-1}^{1}d(\cos\theta)P_4(\cos\theta)[S_3(r,
s, t)-\frac{S_2(r)S_2(t)}{\phi_2}],
\end{equation}
where $t = (r^2 + s^2 - 2 rs \cos\theta)^{1/2}$, and $P_2$ and
$P_4$ are respectively the Legendre polynomials of order two and
four, i.e.,
\begin{equation}
P_2(x) = \frac{1}{2}(3x^2 - 1), \quad P_4(x) =
\frac{1}{8}(35x^4-30x^2+3).
\end{equation}
In our system, the void phase possesses zero elastic moduli, i.e.,
$K_2 = G_2 = 0$, and the moduli of the rock phase are respectively
$K_1 = 4.3$ GPa and $G_2 = 3.8$ GPa. The three-point parameters
$\chi$ and $\eta_2$ are computed by first evaluating the
correlation functions $S_3$ and $S_2$ of the void phase from the
microstructure data and then evaluating the integrals. In the
reported results, the overall Young's modulus of the sandstone is
re-scaled with respect to that of the rock phase.

Similar to the elastic moduli, we use an analytical approximation
\cite{torquato1985effective, jiao2012quantitative} obtained by
truncating the strong-contrast expansion of the effective
diffusion coefficient $D$ after the third order term:
\begin{equation}
\frac{D}{D_2} = \frac{1 + 2\phi_1\beta_{12} - 2\phi_2
\chi\beta_{12}^2}{1- \phi_1\beta_{12}-2\phi_2 \chi \beta_{12}^2},
\end{equation}
where $\phi_1$ and $\phi_2$ are respectively the volume fraction
of the rock and pore phases. The polarization parameter
$\beta_{12}$ is given by
\begin{equation}
\beta_{12} = \frac{D_2 - D_1}{D_2 + 2D_1},
\end{equation}
where $D_1$ and $D_2$ are respectively the diffusivity of the rock
and pore phases. In our sandstone system, the rock diffusivity is
typically negligibly small compared with the pore phase diffusivity, and
thus $D_1/D_2 \approx 0$. The three-point parameter $\chi$ for the
rock phase is given by Eq.~\eqref{chi}, in which the correlation
function $S_3$ and $S_2$ are computed from the microstructural
data for the rock phase (instead of the pore phase as in the case
of elastic moduli). The computed $D$ values are normalized with
respect to $D_2$ in the reported data.

Finally, the fluid permeability is computed using the
approximation~\cite{torquato1990rigorous}
\begin{equation}
k = \frac{2}{3(1-\phi_2)^2}\int_0^\infty[S_2(r)-\phi_2^2]rdr,
\end{equation}
where $\phi_2$ is the volume fraction of the pore phase and $S_2$
is the associated two-point correlation function.

\cutsubsectionup
\subsection{Results}

\begin{figure}[htp]
\centering
\includegraphics[width=16cm]{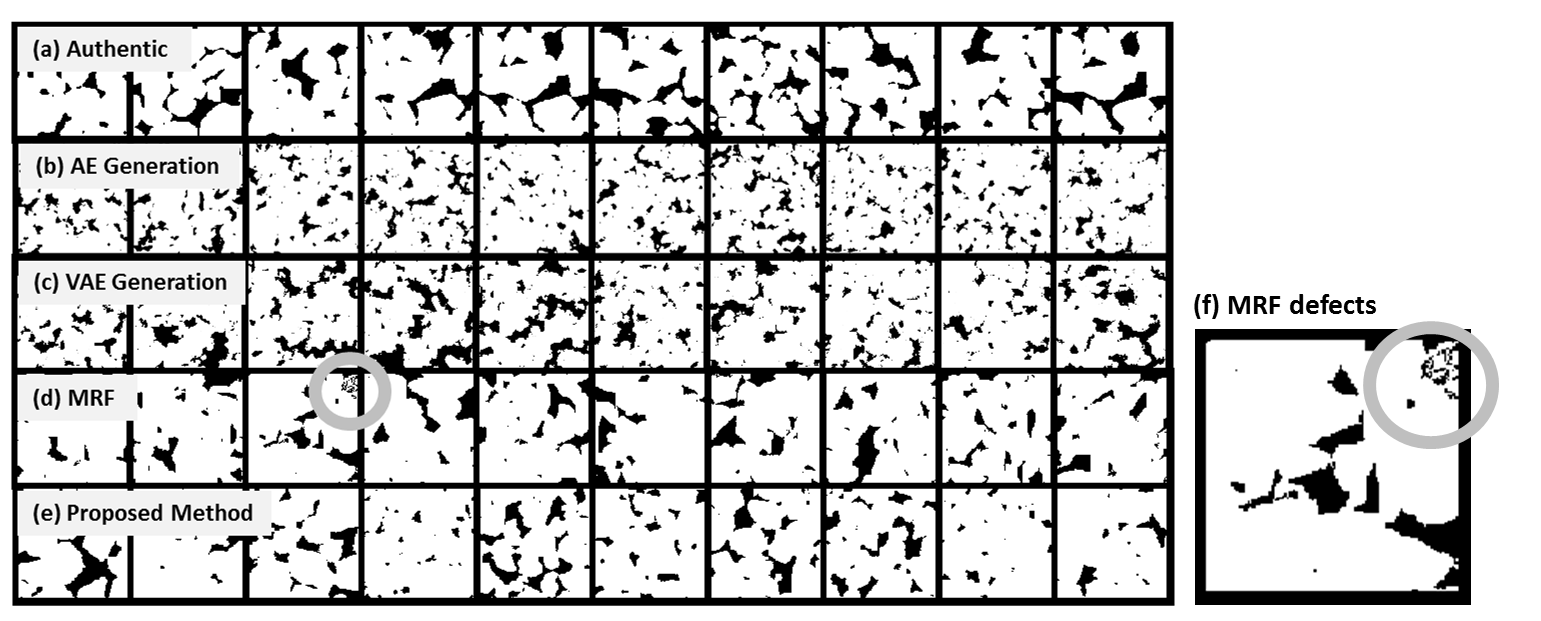}
\caption{(a) The authentic sandstone microstructure samples (b) Generations from a standard autoencoder (c) Generations from a standard VAE (d) Generations from the MRF method in \cite{bostanabad2016stochastic} (e) Generations from the proposed method (f) An enlarged example of MRF generation with artificial defects}
\label{fig:gen_compare}
\end{figure}

\paragraph{The effect of style loss on material generation} 
In Fig.~\ref{fig:gen_compare}, we show that a standard VAE has undesirable generation performance with a small training set (200 samples), and that the incorporation of the style loss can effectively improve the generation quality under the same sample size. Further, the proposed approach also produces more plausible microstructures than the MRF method~\cite{bostanabad2016stochastic}, as one can observe the scattering of unrealistically small particles of the ``stone'' phase from the latter.


\cutparagraphup
\paragraph{Structure-property predictions}
Here we investigate how the prediction accuracy of a structure-property model improves with the additional training data, acquired through the proposed method and the MRF.
To do so, we start by deriving baseline models (for all three structure-property mappings) using 100 data points for training and another 100 for validation. The baseline prediction performance is evaluated using a third set of 100 test data points. 
The true property values are calculated following Sec.~\ref{sec:sandstone}, and normalized for training.

Two sets of artificial microstructures, 1000 samples each, are created using the proposed method and the MRF, respectively. The true properties of these samples are computed. We then use these additional samples to improve the performance of the predictive models. Specifically, an increasing number of additional data points (from 50 to 1000) from the two sets are randomly added to the original training set to update the predictive model. We perform 10 independent random draws for each size of the additional data, and report the means and variances of the resultant test R-squares. The exception is with data size 1000, where all data points are used all together.

\begin{figure}[htp]
\centering
\includegraphics[width=16cm]{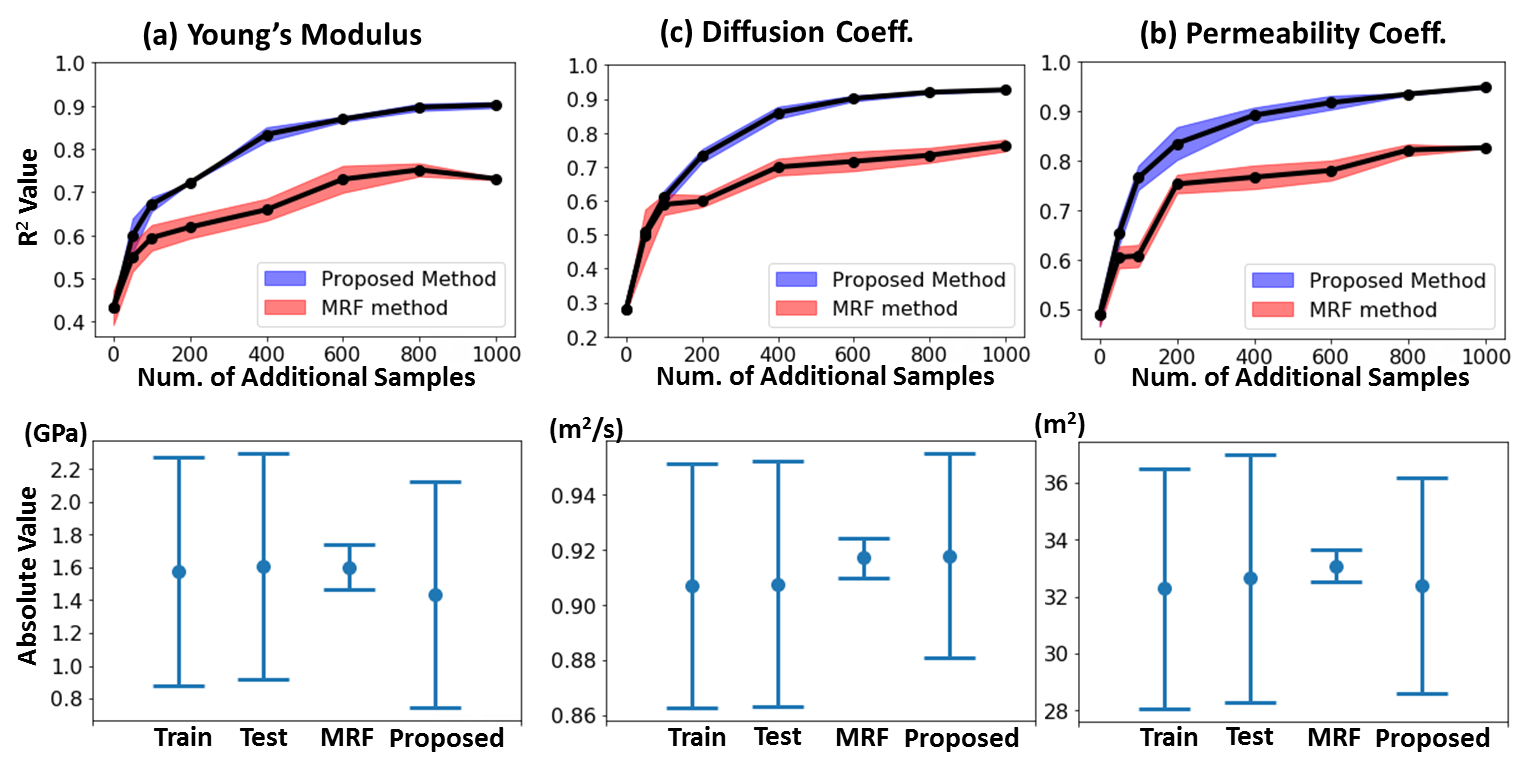}
\caption{(a) Property distributions of microstructures generated from the proposed and the MRF~\cite{bostanabad2016stochastic} models. (b) Test R-square values at an increasing number of additional data points}
\label{fig:results}
\end{figure}

Results are presented in Fig.~\ref{fig:results} and summarized as follows. (1) Microstructures generated from the proposed method have property distributions similar to those of the authentic samples, while those generated from MRF have significantly smaller variances (Fig.~\ref{fig:results}a). While the MRF method ensures variance in the microstructure space, as is evident from Fig.~\ref{fig:gen_compare}, the lack of variance in the volume fraction of its generations may cause the observed small variances in the properties. It is worth noting that while a modified MRF approach that considers variance in volume fraction may improve its generation quality, our approach does not require human identification or design of such features, as variances in morphology are directly learned through the generative model. (2) The statistical difference in properties observed in (1) may explain the significant gap in the prediction performance of the resultant structure-property models, as seen in Fig.~\ref{fig:results}b. 
The narrow distribution of properties of the MRF method also explains the drop in model performance from 800 to 1000 samples: By feeding the network data from one distribution while asking it to predict on those drawn from another, we may encounter deteriorated performance even with an increased training data size.

\section{Discussion}
\label{sec:discussion}
\paragraph{Utilities of the proposed method} The prposed generative model captures low-dimensional representations of data points distributed in high-dimensional spaces.
The model can be used to facilitate more cost-effective design of optimal and feasible microstructures by providing extra data to improve predictive structure-property models. This value could be particularly significant when the bottleneck of the design task is the high acquisition cost of microstructure data. It is also worth noting that the model may also help to reduce the cost of structure-property mappings, through its identification of low-dimensional material representations. For example, during the training of a predictive structure-property model, one may adopt an active learning strategy to only sample microstructures with high uncertainties in their property predictions, thus improving the predictive model cost-effectively. 
The model and the learning algorithm are also generally applicable, as they are independent from any specific material system, except that minor model architecture tuning may be needed to incorporate different morphological complexities.


\subsection{Physics-based generative model}
The successful application of style loss in this paper inspires future investigation on whether physics-based loss can be applied to further improve the quality of microstructure generations through a generative model with limited data. Specifically, instead of using a style loss, we can measure the similarity between input and output microstructures based on their properties. For example, for microstructures that satisfy certain equilibrium conditions (governed by the underlying process-structure mapping), we can train the generative model to minimize the violation of these conditions, and thus enforcing the artificial samples to be physically meaningful.

\section{Conclusions}
\label{sec:conclusions}
This paper proposed a method for generating an arbitrary amount of artificial microstructure samples with low computation cost and a small amount of training samples. The key contribution is the incorporation of a style loss into the training to significantly improve the quality of the artificial microstructures. For verification, we applied the proposed method to generating sandstone samples. Results showed that our method is more data-efficient at improving the prediction performance of a structure-property mapping than a state-of-the-art Markov Random Field method. The findings from this paper inspired future investigation into physically meaningful generative models for accelerating microstructure-mediated material design.



\bibliographystyle{elsarticle-num}

\end{document}